\documentstyle[aps]{revtex}
\begin{document}
\author{Daniel M. Danchev}
\address{Institute of Mechanics, Bulgarian Academy of Sciences, Acad. G. Bonchev Str.
bldg. 4, 1113 Sofia, Bulgaria}
\title{Exact Three Dimensional Casimir Force Amplitude, $C$-function and Binder's
Cumulant Ratio: Spherical Model Results }
\date{\today}
\maketitle

\begin{abstract}
The three dimensional mean spherical model on a hypercubic lattice with a
film geometry $L\times \infty ^2$ under periodic boundary conditions is
considered in the presence of an external magnetic field $H$. The universal
Casimir amplitude $\Delta $ and the Binder's cumulant ratio $B$ are
calculated exactly and found to be $\Delta =-2\zeta (3)/(5\pi )\approx
-0.153051$ and $B=2\pi /(\sqrt{5}\ln ^3[(1+\sqrt{5})/2]).$ A discussion on
the relations between the finite temperature $C$-function, usually defined
for quantum systems, and the excess free energy (due to the finite-size
contributions to the free energy of the system) scaling function is
presented. It is demonstrated that the $C$-function of the model equals $4/5$
at the bulk critical temperature $T_c$. It is analytically shown that the
excess free energy is a monotonically increasing function of the temperature 
$T$ and of the magnetic field $|H|$ in the vicinity of $T_c.$ This property
is supposed to hold for any classical $d$-dimensional $O(n),n>2,$ model with
a film geometry under periodic boundary conditions when $d\leq 3$. An
analytical evidence is also presented to confirm that the Casimir force in
the system is negative both below and in the vicinity of the bulk critical
temperature $T_c.$

\bigskip

\end{abstract}

\section{INTRODUCTION}

The Casimir effect is a phenomenon common to all systems characterized by
fluctuating quantities satisfying some conditions on the boundaries of the
system (for a general review on the Casimir effect see, e.g., \cite{MT97}, 
\cite{K94}). In the statistical mechanical systems the Casimir force is
usually characterized by the excess free energy 
\begin{equation}
f_{{\rm {a,b}}}^{{\rm {ex}}}(T,L)=f_{{\rm {a,b}}}(T,L)-Lf_{{\rm {bulk}}}(T),
\label{fexd}
\end{equation}
due to the finite size contributions to the free energy of finite systems
with a film geometry $L\times \infty ^2,$ where boundary conditions $a$ and $%
b$ are imposed on the surfaces bounding the system across the direction $L$.
Here $f_{{\rm {a,b}}}(T,L)$ is the full free energy per unit area (and per $%
k_BT$ ) of such a system and $f_{{\rm {bulk}}}$ is the corresponding bulk
free energy density. The Casimir force 
\begin{equation}
f_{{\rm {Casimir}}}^{{\rm {a,b}}}(T,L)=-\frac{\partial f_{{\rm {a,b}}}^{{\rm 
{ex}}}(T,L)}{\partial L}  \label{def}
\end{equation}
then arises naturally in the thermodynamics of these confined systems.

For $O(n)$-symmetric model systems ($n\geq 1$ ), depending on the boundary
condition $(a,b)$ and on $n,$ $f_{{\rm {a,b}}}^{{\rm {ex}}}(T,L)$ may or may
not contain contributions independent of $L$. For the Ising-like systems,
i.e. $n=1,$ these can be the surface free energies $f_{{\rm {s,a}}}(T)$ and $%
f_{{\rm {s,b}}}(T)$, and the interface free energy $f_i(T)$ (for brevity we
consider the dependence on the temperature $T$ only). For the $O(n),n\geq 2,$
models these will be only the contributions stemming from the surface free
energies because the analog of the interface free energy is the helicity
modulus $\Upsilon (T)$ and the corresponding contribution is of the order $%
\Upsilon (T)/L$. In general {\it {at}} the critical temperature $T_c$ (of
the corresponding bulk, i.e. $L=\infty $, system) the full free energy $f_{%
{\rm {a,b}}}(T,L)$ has the asymptotic form 
\begin{equation}
f_{{\rm {a,b}}}(T_c,L)\cong Lf_{{\rm {bulk}}}(T_c)+f_{{\rm {s,a}}}(T_c)+f_{%
{\rm {s,b}}}(T_c)+L^{-(d-1)}\Delta _{{\rm {a,b}}}+\cdots ,  \label{fattc}
\end{equation}
where $d$ is the dimensionality of the considered system and $\Delta _{{\rm {%
a,b}}}$ is the so-called Casimir amplitude. The $L$-dependence of the
Casimir term (the last one in Eq. (\ref{fattc})) follows from the scale
invariance of the free energy and has been derived by Fisher and de Gennes 
\cite{FD}. The amplitude $\Delta _{{\rm {a,b}}}$ is {\it {universal}},
depending on the bulk universality class {\it {and}} the universality
classes of the boundary conditions \cite{K94}, \cite{privman}. In the
present article we will only consider the case of periodic boundary
conditions (which implies that $f_{{\rm {s,a}}}=f_{{\rm {s,b}}}=\Upsilon
\equiv 0$ ). Then, according to the standard finite-size scaling theory
(see, e.g., \cite{privman} for a general review), near the critical
temperature $T_c$ and in the presence of a small external magnetic filed $h$
the behavior of $f^{{\rm {ex}}}$ is given by

\begin{equation}
f^{{\rm {ex}}}(t,h,L)=L^{-(d-1)}X^{{\rm {ex}}}(a_ttL^{1/\nu },a_hhL^{\Delta
/\nu }),  \label{fss}
\end{equation}
where $t=(T-T_c)/T_c$ is the reduced temperature, $h=H/(k_BT)$, $a_t$ and $%
a_h$ are nonuniversal scaling factors, $X^{{\rm {ex}}}$ is {\it {universal}}
(usually geometry dependent) scaling function, $X^{{\rm {ex}}}(0,0)\equiv
\Delta _{{\rm {per}}}$, and $\nu $ and $\Delta $ are the corresponding
(universal) scaling exponents.

An interesting point of view on the properties of the excess free energy
comes from the finite-temperature generalizations of the Zamolodchikov's $C$%
-theorem \cite{Z} for quantum systems with arbitrary dimensionality due to
Netto and Fradkin \cite{NF} (see also Zabzin \cite{MZ}; for a general review
on phase transitions in quantum system see, e.g., \cite{Sreview}, \cite{SGCS}%
). They define from the free energy a function $C$ of the coupling constants
and the temperature which is a positive and, in the regimes where the
quantum fluctuations dominate, a monotonically increasing function of the
temperature. The $C$ function is, in fact, an {\it {analog of the excess
free energy}} of the system that they consider.

Before passing to a discussion of some details it seems necessary to comment
on the well known point that for temperature driven phase transitions with $%
T_c>0$ the quantum fluctuations are unimportant near the temperature
critical point. Therefore, it seems that the properties of the system around
one quantum critical point (with respect to a given quantum parameter, say, $%
g$) at $T=0$ cannot tell us anything about the properties of this system
around its temperature critical point $T_c>0.$ In fact the dimensional
crossover rule helps to make a bridge between these phenomena. According to
this rule the critical singularities (with respect to $g,$ $T=0$) of a $d$%
-dimensional quantum system are {\it {formally }} equivalent to those of a $%
d+z$ classical one ($z$ is the dynamical critical exponent) and critical
temperature $T_c>0$. On that idea are actually based the investigations of
the low-temperature effects in quantum systems (see, e.g., \cite{CHN}-\cite
{S}), i.e. one considers an {\it {effective system}} with $d$ infinite space
and $z$ {\it {finite }}``temperature'' (``imaginary-time'') dimensions $%
L_T\sim \left( \hbar /(k_BT)\right) ^{1/z}$ with periodic boundary
conditions, and applies the methods of the finite-size scaling theory (in
what follows we will set $\hbar =k_B=1$). An exact lattice realization of
these ideas is presented in \cite{T98}.

Since the generalizations of the Zamolodchikov's $C$-theorem are formulated
for quantum systems with $z=1,$ in the remainder we will focus our attention
on such class of systems only. For these systems Netto and Fradkin define 
\cite{NF} the dimensionless function 
\begin{equation}
C\left( \beta ,g,a\right) =-\beta ^{d+1}\tilde{n}(d)\lim_{V\rightarrow
\infty }V^{-1}\left[ F_V\left( \beta ,g,a\right) -E_0\left( g,a\right)
\right] ,  \label{Cdef}
\end{equation}
where $E_0$ is the zero-temperature energy, i.e. the energy of the
``infinite'' system, $V$ is the volume ($V\rightarrow \infty ,$ but $N/V$ is
fixed, where $N$ is the number of particles), $\tilde{n}(d)$ is a positive
real quantity, $\beta =1/T$ , $F_V$ is the full free energy of the
``finite'' system (where the only ``finite'' dimension is the
``temperature'' one, i.e. the ``geometry'' of the system is $\infty ^d\times
L_T$) and $a$ is the characteristic length scale of the lattice. The real
positive quantity $\tilde{n}(d)$ is supposed to be of the form $v^d/n(d),$
where $n(d)$ is a positive real number (which depends only on the
dimensionality of the system) and $v$ is the characteristic velocity (e.g.
the velocity of the quasiparticles) in the system. Obviously, the exact
choice of $n(d)$ does not effect the monotonicity properties of the $C$%
-function. In \cite{NF} the definitions $n(d)=\Gamma ((d+1)/2)\zeta
(d+1)/\pi ^{(d+1)/2}$ for bosons and $n(d)=\Gamma ((d+1)/2)\zeta
(d+1)(2-2^{1-d})/\pi ^{(d+1)/2}$ for fermions have been proposed.

In accordance with the dimensional crossover rule the statement that $C$ is
a positive and a monotonically increasing function of the temperature can be
``translated'' in a statement that the function $-X^{{\rm {ex}}}$ of the
corresponding classical system is positive and a monotonically increasing
function of $L^{-1},$ see Eqs. (\ref{fexd}) and (\ref{fss}) (of course, the
last is equivalent to a statement, that $X^{{\rm {ex}}}$ is a negative and a
monotonically increasing function of $L$ ). In \cite{NF}, \cite{MZ} it is
shown that the monotonicity of the $C$-function is related to the absence of
long range order in the systems under consideration. The existence of long
range order destroys the general validity of the monotonicity. Within the
classical systems no long-range order exists above their bulk critical
point. So, we expect the statement formulated for $X^{{\rm {ex}}}$ to be
generally valid above $T_c$ for any classical system. Supposing that this is
true and recalling that in the vicinity of $T_c$ $X^{{\rm {ex}}}$ is a
function of the scaling variables $x_1=a_ttL^{1/\nu }$ and $%
x_2=a_hhL^{\Delta /\nu },$ which both are monotonically increasing functions
of $L,$ we come to the conclusion that {\it {\ in the vicinity of its
critical temperature }} $(T\ge T_c)$ {\it {\ the excess free energy of a
given system is a monotonically increasing function of any of its scaling
parameters when the other one is kept fixed. }} Since $x_1$ and $x_2$ are
monotonically increasing functions of the temperature and the magnetic
field, respectively, the last implies that $X^{{\rm {ex}}}$, {\it {in the
vicinity of }} $T_c$, {\it {is a monotonically increasing function of }}$t$ (%
$t>0$) {\it {and }} $h$ too. It is possible to present some arguments to
support that {\it the above statement can be extended to the region }$t<0$%
{\it \ for }$O(n),n\geq 2,${\it \ systems} in contrast with the Ising-like
systems. The reasoning for the difference in the expected behavior of the
excess free energy in $O(n)$ and Ising-type models is closely related to the
well know differences in the behavior of the correlation length $\xi _\infty
(T)$ in these models: in the Ising model $\xi _\infty (T)<\infty $ both
below and above the bulk critical temperature, whereas in $O(n),$ $n\geq 2,$
models below $T_c$ and in the absence of an external filed ($h=0$), due to
the existence of soft modes in the system (spin waves), $\xi _\infty (T)$ is
identically infinite. On that basis one expects that, away from $T_c,$ $X^{%
{\rm {ex}}}$ will tend to zero exponentially fast in $L$ (see, e.g., \cite
{privman}) for the Ising-type models, and, therefore, being of the order of $%
L^{-(d-1)}$ around $T_c,$ $X^{{\rm {ex}}}$ cannot be a monotonic function of
its scaling parameters in the vicinity of $T_c.$ In $O(n),$ $n\geq 2,$
models the finite size corrections should be essential not only in the
vicinity but also below $T_c$ \cite{danchev}. In other words, we expect the
monotonicity in the behavior of the correlation length in $O(n),$ $n\geq 2,$
models around $T_c$ to be mirrored by a corresponding monotonic behavior of
the excess free energy. If an external field is applied ($h\neq 0$) then $%
\xi _\infty (T,h)<\infty $ and, of course, we expect that $X^{{\rm {ex}}%
}\rightarrow 0$ exponentially fast with $L$ again, similarly to the
Ising-like systems behavior. But, since $X^{{\rm {ex}}}<0,$ for any fixed $%
t<0$ the last implies that $X^{{\rm {ex}}}$ will be a monotonically
increasing function of the magnetic field in the under critical vicinity of $%
T_c$ too.

The statements presented above, should be considered, of course, only as a 
{\it {plausible hypothesis}}, which has to be checked in order to probe the
region of its validity. For example, it is under question if the
monotonicity property of $X^{{\rm {ex}}}$ will still hold if the finite
system undergoes a phase transitions of its own. It is reasonable to believe
that the hypothesis holds for any $O(n),$ $n\geq 2$, system with $d\leq 3$
(then in the finite system with short range interaction there will be no
``real'' phase transition).

In the present article we will show, within the three-dimensional mean
spherical model, that {\it {in the vicinity of }} $T_c$ {\it {the excess
free energy scaling function }} $X^{{\rm {ex}}}$ {\it {is}}, indeed, {\it {a
monotonically increasing function of any of its scaling parameters}} ($x_1$
and $x_2$) {\it {when the other one is kept fixed. }}The last implies that $%
X^{{\rm {ex}}}$ is a monotonically increasing function of $t,h$ and $L$
above $T_c,$ and monotonically increasing, with respect to $t$ and $h,$ but
monotonically decreasing, with respect to $L,$ function below $T_c.$

Let us turn now to the behavior of $f_{{\rm {Casimir}}}.$ From Eqs. (\ref
{def}) and (\ref{fss}) it immediately follows \cite{danchev} 
\begin{equation}
f_{{\rm {Casimir}}}(t,h,L)=L^{-d}X_{{\rm {Casimir}}}(x_1,x_2),  \label{fssc}
\end{equation}
where the Casimir force scaling function is

\begin{equation}
X_{{\rm {Casimir}}}(x_1,x_2)=(d-1)X^{{\rm {ex}}}(x_1,x_2)-\frac 1\nu x_1%
\frac \partial {\partial x_1}X^{{\rm {ex}}}(x_1,x_2)-\frac \Delta \nu x_2%
\frac \partial {\partial x_2}X^{{\rm {ex}}}(x_1,x_2).  \label{casp}
\end{equation}
Note, that $X_{{\rm {Casimir}}}$ is again a {\it {universal}} function of $%
x_1$ and $x_2$. We remind that for finite-size systems this means that $X_{%
{\rm {Casimir}}}$ will be the same for all systems of the same universality
class {\it {and }}geometry and boundary conditions. It is believed that if $%
a\equiv b$ the Casimir force will be negative (see, e.g., \cite{evst}, \cite
{PE}; strictly speaking, for an Ising-like system this is supposed to be
true above the wetting transition temperature $T_w$ \cite{evst}, \cite{PE}, 
\cite{dietrich}). In the case of a fluid confined between identical walls
this implies that then the net force between the plates will be attractive
for large separations. One of the goals of the present article is to prove
analytically this general expectation, i.e. that $X_{{\rm {Casimir}}%
}(x_1,x_2)<0$ for any $(x_1,x_2)\in {\bf {R}}^2,$ on the example of one
exactly solvable model. We will also show that if $T<T_c$ and $H=0$ the
Casimir force is a monotonically increasing function of the temperature. We
believe that these properties are valid for any $O(n),n\geq 2$ models.

The full temperature dependence of the Casimir force has been investigated
exactly in two-dimensional Ising strips by Evans and Stecki \cite{evst},
whereas the upper critical temperature dependence of the force in $O(n)$
systems has been considered by Krech and Dietrich \cite{krdi} by means of
the field-theoretical renormalization group theory in $4-\epsilon $
dimensions. (For the Ising-like case they have derived also some results for 
$T<T_c$.) The only example where exact expression for the Casimir force as a
function of both the temperature and the magnetic field is available is that
one of the three-dimensional mean spherical model \cite{danchev}. By
numerical evaluation of the expressions derived there it has been shown that
the force is negative, i.e. it is consistent with an attraction of the
plates confining the system. The most results available at the moment are
for the Casimir amplitudes $\Delta _{a,b}$. For two-dimensional systems at $%
T=T_c$ by using conformal-field theory methods the amplitudes are exactly
known for a large class of two-dimensional models \cite{K94}, \cite{Cardy}, 
\cite{PHA91}. In addition to the ``flat geometries'' recently some results
about the Casimir amplitudes between spherical particles in a critical fluid
have been derived too \cite{EBR}. For $d=3$ the results for the Casimir
amplitudes available in the Ising-like case have been obtained by
Migdal-Kadanoff renormalization-group calculations \cite{I}, by some
interpolation of the exact values for $d=2$ and $d=4$ \cite{krdi}, and,
relatively recently, by Monte Carlo methods \cite{kl96}, \cite{k97}. For $%
n\geq 2$ the only existing results are obtained by the $\epsilon $-expansion
technique, where the calculations are performed up to the first order in $%
\epsilon $ \cite{krdi}.

In the present article the hypotheses for the monotonicity of the excess
free energy and that the Casimir force is negative under periodic boundary
conditions will be verified analytically on the example of the
three-dimensional mean spherical model. We will present also simple
analytical results for the universal values of the Casimir amplitude and the
Binder's cumulant ratio. If one takes the normalization factor of the analog
of the $C$-function in the form for bosons (this will keep the $C$-function
of the critical Gaussian model to be $C=1$ for any $d$), it will be shown
that the {\it {``}}$C${\it {-function of the three dimensional spherical
model'' is }} $4/5$ {\it {\ at the critical point}}. As it is well known,
the infinite translational invariant spherical model is equivalent to the $%
n\rightarrow \infty $ limit of the corresponding $n$-component system \cite
{SKT}.

The results we are going to present are an extension and continuation of
those published in \cite{danchev}. In the notations and the definitions in
the remainder we will closely follow \cite{danchev}. That is why here we
only briefly recall, in Section 2, the definition of the model and give the
final expressions, obtained there, for the excess free energy and the
Casimir force which will be our starting expressions for the aims of the
current article. In Section 3 we verify the hypotheses, formulated above,
for the excess free energy and the Casimir force. In Section 4 we derive the
exact universal values for the Casimir amplitude and the Binder's cumulant
ratio. The paper closes with concluding remarks given in Section 5.

\section{THE\ MODEL}

We consider the ferromagnetic mean-spherical model (see, e.g., \cite{BF}, 
\cite{R} for a general review) on a fully finite $d$-dimensional hypercubic
lattice $\Lambda _d$ of ${}\mid \!\Lambda \!\mid $ sites and with block
geometry $L_1\times L_2\times \cdots \times L_d$, where $L_i,i=1,\cdots ,d$
are measured in units of the lattice spacing. The Hamiltonian has the form{}

\begin{equation}
\beta {\cal {H}}_{\Lambda} ^{b.c.}\left( \left\{ \sigma _i \right\} _{i\in
\Lambda }\right) =-\frac 12K\sum_{i,j\in \Lambda }J_{ij}^{b.c.}\sigma
_i\sigma _j+s\sum_{i\in \Lambda }\sigma _i^2-h\sum_{i\in \Lambda }\sigma _i.
\label{H}
\end{equation}

Here $\sigma _i\in I\!\!R,i\in \Lambda _d$ ($\sigma _i\equiv \sigma \left( 
{\bf {r}}_i\right) $) is a variable, describing the spin on lattice site $i$
(at ${\bf {r}}_i$), $s$ is the spherical field, $K$ is a dimensionless
coupling, $J_{ij}^{b.c.}$ is a matrix with dimensionless elements, so that $%
\left( K/\beta \right) J_{ij}^{b.c.}$ is the exchange energy between the
nearest neighbors (under boundary conditions $b.c.$) spins at sites $i$ and $%
j$ (of course, $J_{ij}^{b.c.}=J_{ji}^{b.c.}$), and $h$ is the external
magnetic field. The dependence on the boundary conditions is denoted by a
superscript ${b.c.}$

The scaling function of the free energy density of the spherical model has
been discussed in details in the literature for different boundary
conditions, dimensionalities and geometries of the system, for both the
cases of short as well as for long range interactions in the Hamiltonian 
\cite{privman}, \cite{R}, \cite{Pathria}, \cite{history}, \cite{DB}. By any
of the approaches used there one can, of course, derive an expression for
the excess free energy scaling function. Here, for $d=3$ and under periodic
boundary conditions we will take it in the form given in \cite{danchev} 
\begin{eqnarray}
X^{{\rm {ex}}}(x_1,x_2) &=&\frac 12\left( 4\pi \right) ^{-3/2}\left[
\sum_{k=0}^\infty \frac{\left( -1\right) ^k\left( y_L^{k+1}-y_\infty
^{k+1}\right) }{\left( k+1\right) !\left( k-1/2\right) }\right.  \label{xec}
\\
&&-\sqrt{4\pi }\int_1^\infty dxx^{-2}\left( 1+2R(4\pi ^2x)\right) \exp
[-y_Lx]  \nonumber \\
&&-2\int_0^1dxx^{-5/2}R(1/4x)\exp [-y_Lx]+\int_1^\infty dxx^{-5/2}\exp
\left( -y_\infty x\right) \Biggr]  \nonumber \\
&&+\frac 12x_2^2\left( \frac 1{y_\infty }-\frac 1{y_L}\right) +\frac 12%
x_1\left( y_\infty -y_L\right) ,  \nonumber
\end{eqnarray}
where

\begin{equation}
R\left( x\right) =\sum_{q=1}^\infty \exp \left[ -xq^2\right],  \label{R}
\end{equation}
\begin{equation}
x_1=\left( K-K_c\right) L,x_2=K_c^{-1/2}hL^{5/2}  \label{sv}
\end{equation}
are the scaling variables (note the difference in the definitions of $x_1$
here and in the Introduction; now $x_1$ decreases when $T$ increases),

\begin{equation}
K_c=\int_0^\infty dx\left[ \exp \left( -2x\right) I_0\left( 2x\right)
\right] ^3=0.25273\ldots  \label{KK}
\end{equation}
is the critical coupling, and $y_L$ and $y_\infty $ are the solutions of the
spherical field equations that follow from (\ref{xec}) by requiring the
first partial derivatives of the right-hand side of (\ref{xec}) with respect
to $y_L$ and $y_\infty $ to be zero.

For the finite-size scaling function of the Casimir force one immediately
obtains from Eqs. (\ref{fssc}), (\ref{casp}), (\ref{xec}) and the
definitions of the scaling variables $x_1$ and $x_2$ \cite{danchev} 
\begin{equation}
X_{{\rm {Casimir}}}(x_1,x_2)=2X^{{\rm {ex}}}(x_1,x_2)-\frac 52x_2^2\left( 
\frac 1{y_\infty }-\frac 1{y_L}\right) -\frac 12x_1\left( y_\infty
-y_L\right) .  \label{xcas}
\end{equation}

The equations (\ref{xec}) -- (\ref{xcas}) provide the basis of our further
analysis.

\section{VERIFICATION OF THE HYPOTHESES}

We will prove analytically that the finite-size scaling function of the
excess free energy, given by Eq. (\ref{xec}), is a monotonically increasing
function of any of its scaling parameters $x_1$ and $x_2$ when the other one
is kept fixed. First, by using the identity 
\begin{equation}
\sum_{k=0}^\infty \frac{\left( -1\right) ^ky^{k+1}}{\left( k+1\right)
!\left( k-1/2\right) }=-\frac{4\sqrt{\pi }}3y^{3/2}-\frac 23y\int_1^\infty
x^{-3/2}\exp \left( -x\right) dx-\frac 23(1-\exp (-y)),  \label{i1}
\end{equation}
the Jacobi identity for the $R$ function (see Eq. (\ref{R})) 
\begin{equation}
R\left( 4\pi ^2x\right) =\frac 12\left\{ \frac 1{\sqrt{4\pi x}}\left[
1+2R\left( \frac 1{4x}\right) \right] -1\right\} ,  \label{iR}
\end{equation}
and taking into account that 
\begin{equation}
\int_0^\infty \frac{dx}{x^{5/2}}R\left( \frac 1{4x}\right) \exp \left(
-xy\right) =4\sqrt{\pi }\left[ \sqrt{y}\;{\rm {Li}_2}\left( \exp \left( -%
\sqrt{y}\right) \right) +{\rm {Li}_3}\left( \exp \left( -\sqrt{y}\right)
\right) \right] ,  \label{integ}
\end{equation}
after some elementary manipulations we obtain from (\ref{xec}) 
\begin{eqnarray}
X^{{\rm {ex}}}(x_1,x_2) &=&-\frac 1{2\pi }\left[ \frac 16\left(
y_L^{3/2}-y_\infty ^{3/2}\right) +\sqrt{y_L}\;{\rm {Li}_2}\left( \exp \left(
-\sqrt{y_L}\right) \right) +{\rm {Li}_3}\left( \exp \left( -\sqrt{y_L}%
\right) \right) \right]  \nonumber \\
&&+\frac 12x_2^2\left( \frac 1{y_\infty }-\frac 1{y_L}\right) +\frac 12%
x_1\left( y_\infty -y_L\right) ,  \label{xecff}
\end{eqnarray}
where ${\rm {Li_p}}(z)$ are the polylogarithm functions. The main advantage
of the above representation of $X^{{\rm {ex}}}$ is the existence of some
nontrivial identities \cite{L81}, \cite{S93} for the polylogarithm functions
(see next Section) that allow the universal constant $\Delta =X^{{\rm {ex}}%
}(0,0)$ to be expressed in a simple closed form.

The spherical field equations for $y_L$ and $y_\infty $ can be now rewritten
in the well known and very simple forms (see, e.g., for $h=0$, Eq. (86) in 
\cite{Pathria}) 
\begin{equation}
x_1=\frac{x_2^2}{y_L^2}-\frac 1{2\pi }\ln \left[ 2\sinh \left( \frac 12\sqrt{%
y_L}\right) \right] ,  \label{eqff}
\end{equation}
and 
\begin{equation}
x_1=\frac{x_2^2}{y_\infty ^2}-\frac 1{4\pi }\sqrt{y_\infty }\;,  \label{eqif}
\end{equation}
where the first equation is for the finite and the second one for the
infinite system, respectively. In order to obtain Eq. (\ref{eqff}) use has
been made of the facts that $d{\rm {Li}_p}(x)/dx={\rm {Li}_{p-1}}(x)/x$ and $%
{\rm {Li}_1}(x)=-\ln (1-x).$ Let us denote by $g_L(x_2,y_L)$ the right-hand
side of Eq. (\ref{eqff}) and by $g_\infty (x_2,y_\infty )$ the right-hand
side of Eq. (\ref{eqif}). Then, it is easy to see, that 
\begin{equation}
g_L(x_2,y)=g_\infty (x_2,y)-\frac 1{2\pi }\ln \left[ 1-\exp \left( -\sqrt{y}%
\right) \right] .  \label{g}
\end{equation}
From the above equation and having in mind that in Eqs. (\ref{eqff}) and (%
\ref{eqif}) $y_L>0,$ $y\geq 0$ we conclude that 
\begin{equation}
g_L(x_2,y)>g_\infty (x_2,y).  \label{uneq}
\end{equation}
It is also elementary to verify that $g_L(x_2,y_L)$ and $g_\infty
(x_2,y_\infty )$ are monotonically decreasing functions of $y_L$ and $%
y_\infty ,$ respectively. Let now $y_\infty (x_1,x_2)$ be the solution of
Eq. (\ref{eqif}) for given $x_1$ and $x_2.$ Then, from Eq. (\ref{uneq}), the
fact that $g_L(x_2,y)$ is a monotonically decreasing function of $y,$ and
that for the solution $y_L(x_1,x_2)$ of Eq. (\ref{eqff}) one should have $%
g_L(x_2,y_L)=g_\infty (x_2,y_\infty ),$ we obtain 
\begin{equation}
y_L(x_1,x_2)>y_\infty (x_1,x_2).  \label{uneqy}
\end{equation}
We are now ready to deal with the monotonicity properties of the excess free
energy scaling function. From (\ref{xecff}) and having in mind the spherical
filed equations (\ref{eqff}) and (\ref{eqif}) we derive 
\begin{equation}
\frac{\partial X^{{\rm {ex}}}}{\partial x_1}=-\frac 12(y_L-y_\infty ),
\label{mt}
\end{equation}
and 
\begin{equation}
\frac{\partial X^{{\rm {ex}}}}{\partial x_2}=-x_2\left( \frac 1{y_L}-\frac 1{%
y_\infty }\right) .  \label{mf}
\end{equation}
From these expressions and Eq. (\ref{uneqy}), taking into account the
definitions of the scaling variables (\ref{sv}), we obtain that the excess
free energy scaling function is a monotonically increasing function of both
the temperature $T$ and the magnetic field $|H|$. As a function of the
finite size $L$ of the system the scaling function is monotonically
increasing above and decreasing below $T_c.$ These properties of the scaling
function as a function of the scaling variables $x_1$ and $x_2$ are
illustrated in Fig 1. One clearly sees that for any fixed $x_2$ the scaling
function is a monotonically decreasing function of $x_1,$ and, for any fixed 
$x_1$ a monotonically increasing function of $|x_2|.$ Finally, it is worth
mentioning, that, for $x_2=0$ from (\ref{xecff}) and (\ref{uneqy}) it
immediately follows that $X^{{\rm {ex}}}<0.$ From Fig. 1 one observes that
this is true also for $x_2\neq 0.$

We turn now to properties of the Casimir force. Our aim is to show that the
force is negative under periodic boundary conditions for any values of $T$
and $H.$ The finite-size behavior of the Casimir force in the vicinity of
the critical point is given by Eq. (\ref{fssc}) where the scaling function
is given by Eq. (\ref{xcas}). For $T<T_c$ the same expressions are actually
valid with the only difference that the definition of the variable $x_2$ now
should be $x_2=K^{-1/2}hL^{5/2}$ and $x_1\gg 1.$ Here we are not going to
discuss if then the above expressions can be simplified further, e.g., being
a function of a given combination of $x_1$ and $x_2,$ as it is usually the
case of first order phase transitions \cite{FP}). For $T\ $away above $T_c$
the Casimir force, as it has been shown in \cite{danchev}, tends to zero
exponentially fast with $L$ in a full accordance with the general
expectations about its behavior above the critical point. We will not be
interested in the explicit form of these exponentially small corrections.
Having in mind all these comments, for the behavior of the Casimir force for
any $T$ and $H$ one obtains explicitly 
\begin{eqnarray}
f_{{\rm {Casimir}}}(t,h,L) &=&L^{-3}\left\{ \frac 32x_2^2\left( \frac 1{y_L}-%
\frac 1{y_\infty }\right) -\frac 12x_1\left( y_L-y_\infty \right) \right. 
\label{fcas} \\
&&\left. -\frac 1\pi \left[ \frac 16\left( y_L^{3/2}-y_\infty ^{3/2}\right) +%
\sqrt{y_L}\;{\rm {Li}_2}\left( \exp \left( -\sqrt{y_L}\right) \right) +{\rm {%
Li}_3}\left( \exp \left( -\sqrt{y_L}\right) \right) \right] \right\} . 
\nonumber
\end{eqnarray}
Since the inequality (\ref{uneqy}) is still valid, from the above expression
it immediately follows that $f_{{\rm {Casimir}}}(t,h)<0.$ Numerical
evaluation of the behavior of the finite-size scaling function of the
Casimir force has been given in \cite{danchev}. It is in full agreement with
our analytical result. Finally we show, that for $T<T_c$ and $h=0,$ i.e. $%
x_1>0$ and $x_2=0,$ the Casimir force is a monotonically increasing function
of the temperature, i.e. monotonically decreasing function of $x_1.$ From (%
\ref{fcas}) and taking into account that $y_\infty =0$ when $T<T_c$ we
obtain 
\begin{equation}
\frac d{dx_1}X_{{\rm {Casimir}}}(x_1,0)=-\frac 12y_L+\frac 12x_1\frac{dy_L}{%
dx_1}.  \label{casuneq}
\end{equation}
From (\ref{eqff}) it is easy to see that $dy_L/dx_1<0,$ and, therefore $dX_{%
{\rm {Casimir}}}(x_1,0)/dx_1<0,$ i.e. the Casimir force is an increasing
function of $T$ for $T<T_c$ and $h=0.$

In this way we have completely verified the hypotheses formulated in the
introductory part of the article for the behavior of the excess free energy
and the Casimir force.

\section{CASIMIR AMPLITUDE, $C$-FUNCTION AND BINDER'S CUMULANT RATIO}

Here we will be interested in the properties of the system at its bulk
critical point. This implies $x_1=$ $x_2=0$ with a solution of the spherical
field equations (see Eqs. (\ref{eqif}) and (\ref{eqff})) $y_\infty =0$ and $%
y_L\equiv y_{L,c}=4\ln ^2\left[ \left( 1+\sqrt{5}\right) /2\right] $ (this
value of $y_{L,c}$ is well known and, seems, has been derived for the first
time in \cite{P2}). The problem of determination of the Casimir amplitude
reduces now to exact evaluation of the expression 
\begin{equation}
X^{{\rm {ex}}}(0,0)=-\frac 1{2\pi }\left[ \frac 16y_{L,c}^{3/2}+\sqrt{y_{L,c}%
}\;{\rm {Li}_2}\left( \exp \left( -\sqrt{y_{L,c}}\right) \right) +{\rm {Li}_3%
}\left( \exp \left( -\sqrt{y_{L,c}}\right) \right) \right] .  \label{casam}
\end{equation}
Denoting by $\tau $ the ``golden mean'', i.e. $\tau =\left( 1+\sqrt{5}%
\right) /2,$ it is easy to show that 
\begin{equation}
\exp \left( -\sqrt{y_{L,c}}\right) =\tau ^{-2}=2-\tau ,  \label{eqc}
\end{equation}
which reduces the above problem for $X^{{\rm {ex}}}(0,0)$ to the problem for
evaluation of the expression 
\begin{equation}
a={\rm {Li}_3}\left( 2-\tau \right) -\ln \left( 2-\tau \right) {\rm {Li}_2}%
\left( 2-\tau \right) -\frac 16\ln ^3\left( 2-\tau \right) .  \label{ct}
\end{equation}
Fortunately, this is exactly the problem solved by Sachdev \cite{S93}
studying his example of a conformal field theory in three dimensions. By the
help of some polylogarithm identities he has shown that $a=4\zeta (3)/5.$
Therefore, we obtain for the Casimir amplitude of the three dimensional
spherical model under periodic boundary conditions 
\begin{equation}
\Delta =-\frac{2\zeta (3)}{5\pi }\approx -0.153051.  \label{delta}
\end{equation}
The numerical value of this amplitude has already been reported in \cite
{danchev}. Recalling now that $-X^{{\rm {ex}}}(0,0)$ corresponds to the
analog of the $C$-function for our model and taking the normalization factor
in the form that will keep the $C$-function of the critical Gaussian model
to be $C=1$ for any $d$ (i.e. by taking the normalization in the form for
bosons) we conclude that the {\it {``}} $C${\it {-function of the spherical
model'' is }}$4/5$ (at $T=T_c$ for $d=3$ under periodic boundary conditions).

Let us turn now to determination of the Binder's cumulant ratio for the
considered model. We will use for it the definition of the form \cite
{privman} (up to a prefactor $1/3)$%
\begin{equation}
B_L=-L^{-d}\frac{\chi ^{(4)}(t,h=0,L)}{3\chi ^{(2)}(t,h=0,L)},  \label{bcr}
\end{equation}
where $\chi ^{(n)}$ means the $n$-th derivative with respect of $h$ of the
free energy density at $h=0$ (of course, $\chi ^{(2)}=-\chi ,$ where $\chi $
is the susceptibility of the system). In the vicinity of the critical point
this expression can be rewritten in the form 
\begin{equation}
B_L(x_1)=-\frac 13\left\{ \frac{\partial ^4X(x_1,x_2)/\partial x_2^4}{\left[
\partial ^2X(x_1,x_2)/\partial x_2^2\right] ^2}\right\} _{x_2=0},
\label{binder}
\end{equation}
where $X(x_1,x_2)$ is the finite-size scaling function of the free energy
density. The exact form of this function follows from (\ref{xecff}) just by
omitting the terms depending on $y_\infty $ in it, i.e.

\begin{eqnarray}
X(x_1,x_2) &=&-\frac 1{2\pi }\left[ \frac 16y_L^{3/2}+\sqrt{y_L}\;{\rm {Li}_2%
}\left( \exp \left( -\sqrt{y_L}\right) \right) +{\rm {Li}_3}\left( \exp
\left( -\sqrt{y_L}\right) \right) \right]  \nonumber \\
&&-\frac 12\frac{x_2^2}{y_L}-\frac 12x_1y_L.  \label{xsf}
\end{eqnarray}
From the above expression at the critical point it immediately follows that 
\begin{equation}
B\equiv B_L(x_1=0)=-\left[ 2y_{L,c}^{-1}\left( \frac{\partial y_L}{\partial
x_2}\right) _{x_1=x_2=0}^2-\left( \frac{\partial ^2y_L}{\partial x_2^2}%
\right) _{x_1=x_2=0}\right] .  \label{brc}
\end{equation}
By subsequent differentiation of the spherical filed equation for the finite
system (\ref{eqff}) it is easy to show that at the critical point $\partial
y_L/\partial x_2=0,$ whereas 
\begin{equation}
\left( \frac{\partial ^2y_L}{\partial x_2^2}\right) _{x_1=x_2=0}=\frac{16\pi 
}{y_{L,c}^{3/2}\coth \left( \sqrt{y_{L,c}}/2\right) }.  \label{sd}
\end{equation}
Combining these results and having in mind that $y_{L,c}=4\ln ^2\tau $ we
obtain for the Binder's cumulant ratio at the critical point 
\begin{equation}
B=\frac{2\pi }{\sqrt{5}\ln ^3\tau }\approx 25.21657.  \label{bv}
\end{equation}
Having the exact solution for the spherical field equation and such a simple
form for the free energy density, one can easily determine in an exact
manner the behavior of other physically interesting quantities at $T=T_c.$
For example it is easy to show that the specific heat is of the form 
\begin{equation}
c_L(T_c)=\frac 12-L^{-1}\frac{16\pi }{\sqrt{5}}K_c^2\ln \tau ,  \label{sh}
\end{equation}
and that the critical finite-size correlation length is ($\xi _L=L/\sqrt{y_L}
$ \cite{DB}, \cite{SP86}) 
\begin{equation}
\xi _L(T_c)=\frac 1{2\ln \tau }L  \label{cl}
\end{equation}
(for explicit results of the behavior of $\xi _L$ under other geometries,
boundary conditions and long-rangines of the spin-spin interactions see \cite
{DB}, \cite{SP86}, \cite{AP93}, \cite{history2}).

\section{CONCLUDING REMARKS}

In the present paper we present a hypothesis that in the vicinity of the
bulk critical temperature $T_c$ of $O(n),n\geq 2$, systems with a film
geometry $L\times \infty ^{d-1}$ the excess free energy (due to the finite
size of the system) will be, under periodic boundary conditions, a
monotonically increasing function of the temperature and the magnetic field
if the finite system does not undergo a real phase transition of its own
(i.e. when $d\leq 3$ for systems with short-range interactions). As a
function of the finite size $L$ of the system the finite size scaling
function of the excess free energy is expected to be monotonically
increasing above and decreasing below $T_c.$ This hypothesis, together with
the hypothesis that the Casimir force should be negative under periodic
boundary conditions have been verified {\it {analytically}} on the example
of the three-dimensional mean spherical model. It has been shown that the
force is negative in the whole region of the thermodynamic parameters. In
addition the universal Casimir amplitude $\Delta _{{\rm {per}}}$ and the
Binder's cumulant ratio have been determined exactly in a simple close form
and found to be $\Delta _{{\rm {per}}}=-2\zeta (3)/(5\pi )\approx -0.153051$
and $B=2\pi /(\sqrt{5}\ln ^3[(1+\sqrt{5})/2])\approx 25.21657.$ For
comparison we give the corresponding result for the Ising universality
class, $\Delta _{{\rm {per}}}=-0.1526\pm 0.0010$ \cite{k97}, and $B=$ $%
0.615\pm 0.003$ \cite{BD}, \cite{BD2} obtained by Monte Carlo calculations.
As we see, the value for the Casimir amplitude for the spherical model is 
{\it {surprisingly close }}(within the error bar) to the value reported
above for the Ising model. The vast difference for the cumulant ratio
indicates the lack of a real phase transition in the three dimensional
spherical model film in comparison with the Ising like films. Actually, in
three-dimensional Ising films the situation is more complicated \cite{B}. If
the thickness of the film $L$ is held constant and the other two linear
dimensions $D$ tend to infinity, the cumulant ratio converges to the
two-dimensional Ising value ($B=$ $0.615$). However, if the ratio $L/D$ is
not too small, there exist crossover problems. In any case the value of $B$
is between that one for the two-dimensional system and that one for the
three-dimensional system ($B=0.47$ \cite{FL}). The value of $B$ for the
spherical model shows that the probability distribution at $T_c$ of the
order parameter density is too different from a single Gaussian, where $B=0,$
or from a normalized sum of two Gaussians, where $B=2/3.$ This, of course,
rises the question what then that distribution is, but this question is out
of the scope of the current article. The situation reminds the one of Ising
strips (no real phase transition in the system) with $B=2.46044\pm 0.00006$ 
\cite{privman}, \cite{BD}, \cite{BSD}. The crossover problems in Binder's
cumulant ratio can be studied within the spherical model, considering a $%
3+\varepsilon $ dimensional film, $\varepsilon >0$ (then in the finite
system there will be a real phase transition). This is also an interesting
problem, especially if one takes into account that there are almost no exact
results for the Binder's cumulant ratio, but it is again out of the scope of
the current article.

The results reported in the current investigation are in full agreement with
the predictions of the finite-size scaling theory. Eqs.(\ref{xecff}), (\ref
{eqff}), (\ref{eqif}), (\ref{fcas}) and (\ref{xsf}) give the {\it {universal 
}} finite-size scaling function of the excess free energy, Casimir force and
free energy density. It should be, however, emphasized that in contrast to
the Ising-like case the excess free energy, and, therefore, the Casimir
force in the absence of an external field tend to zero below $T_c$ not in an
exponential in $L$ way. For example, the finite-size scaling functions of
the excess free energy and Casimir force tend to a constant below $T_c$ (see
Eq. (31) in \cite{danchev}). The explanation of this behavior, which, we
believe, is common for all $O(n),n\geq 2$ models, is based on the fact that
due to the existence of soft modes in the system (spin waves) below $T_c$
and in the absence of an external field ($h=0$) $\xi _b$ is identically
infinite. If an external field is applied ($h\neq 0$) then $\xi _b<\infty $,
and, of course, $f^{{\rm {ex}}}\rightarrow 0$ again exponentially fast in $L$%
.

Finally, it is worth mentioning the close parallel that exists between the
properties of the $C-$function defined by Netto and Fradkin \cite{NF}, see
also Zabzin \cite{MZ}, for a $d$-dimensional quantum system as a function of
the temperature $T$ and the properties of the excess free energy scaling
function $-X^{{\rm {ex}}}$ of the corresponding classical system as a
function of $L^{-1}.$ If in the finite system a real phase transition does
not exist, and if the system is somehow equivalent to the $O(n),n>2$ system
we have proposed some arguments that $-X^{{\rm {ex}}}$ is a monotonically
increasing as a function of $L^{-1}$ above $T_c$ and decreasing below $T_c.$
We would expect the same to be true for the $C$-function of the
corresponding quantum system as a function of $T$ around its quantum
critical point. If the classical system is equivalent to some Ising type
model, the same type of arguments we have used for the $O(n),n>2$ models,
taking into account the lack of monotonicity of the correlation length in
the vicinity of $T_c,$ lead to the hypothesis that $-X^{{\rm {ex}}}$ will be
a monotonic function of $L^{-1}$ both below and above $T_c.$ For the
corresponding $C$-function of a quantum system that has its mapping into a
classical Ising system (according to the dimensional crossover rule) this
means that $C$ is a monotonically increasing function of the temperature
both below and above its quantum critical point. This is indeed the case
plotted in Fig. 2 in \cite{NF} for the quantum version of the
two-dimensional Ising model. At the very end, we would like to stress that
the relatively simple picture described here should probably change
significantly, if the finite system undergoes a phase transition of its own.
In that case the upper critical part of the excess free energy scaling
function for $4-\varepsilon $ Ising model is known \cite{krdi} (up to a
first order in $\varepsilon ,\varepsilon >0$). It shows a {\it {minimum}} in 
$X^{{\rm {ex}}},$ as a function of $T$ {\it {slightly above}} $T_c.$
Unfortunately, no results are available for $X^{{\rm {ex}}}$ when $%
T<T_{c,L}, $ where $T_{c,L}$ is the shifted critical temperature of the
finite system. It is possible to investigate the above problems exactly
within the spherical model with $3+\varepsilon $ infinite dimensions. We
hope to return to this problem later.

\section{ACKNOWLEDGMENTS}

The author thanks Prof. K. Binder and Prof. D. P. Landau for clarifying the
situation with the cumulant ratios for $3d$ Ising films, as well as Prof. N.
Tonchev, Prof. J. Brankov and Dr. M. Krech for stimulating and fruitful
discussions.

This work is partially supported by the Bulgarian National Foundation for
Scientific Research, grant MM-603/96.

\newpage\ 

\begin{center}
{\bf {Figure captions} }
\end{center}

\noindent FIG.1. The universal finite-size scaling function of the excess
free energy $X^{{\rm {ex}}}$ as a function of the scaling variables $%
x_1=L(K-K_c)/K_c$ and $x_2=K_c^{-1/2}hL^{5/2}$.{\small \ }For a better
visualization of the properties of $X^{{\rm {ex}}}$ we have allowed $h$ to
change its sign. Of course, $X^{{\rm {ex}}}$ is a symmetric function of $%
x_2. $


\newpage

\begin{references}
\bibitem{MT97}  V. M. Mostepanenko and N. N. Trunov, {\it {The Casimir
Effect and its Applications}} (Clarendon Press, New York, 1997).

\bibitem{K94}  M. Krech, {\it {The Casimir Effect in Critical Systems }}%
(World Scientific, Singapore, 1994).

\bibitem{privman}  V. Privman, in {\it {Finite Size Scaling and Numerical
Simulations of Statistical Systems}}, edited by V. Privman (World
Scientific, Singapore, 1990).

\bibitem{FD}  M. E. Fisher, P. G. de Gennes, C. R. Acad. Sci. Paris B {\bf {%
287}}, 207 (1978).

\bibitem{danchev}  D. Danchev, Phys. Rev. E{\it {\ }} {\bf {53}}, 2104
(1996).

\bibitem{Z}  A. B. Zamolodchikov, JETF Lett. {\bf {43}}, 731 (1986); Sov. J.
Nucl. Phys. {\bf {46}}, 1090 (1987).

\bibitem{NF}  A. H. Castro Neto and E. Fradkin, Nuclear Physics B {\bf {400}}%
[FS], 525 (1993).

\bibitem{MZ}  M. Zabzin, hep-th/9705015.

\bibitem{Sreview}  S. Sachdev, in {\it {Strongly Correlated Magnetic and
Superconducting Systems}}, edited by G. Sierra and M. A. Martin-Delgado,
(Springer, Berlin, 1996).

\bibitem{SGCS}  S. L. Sondhi, S. M. Girvin, J. P. Carini and D. Shahar, Rev.
Mod. Phys. {\bf {69}}, 315 (1997).

\bibitem{CHN}  S. Chakravarty, B. I. Halperin, and D. R. Nelson, Phys. Rev. B%
{\it {\ }}{\bf {39}}, 2344 (1989).

\bibitem{SY}  S. Sachdev and J. Ye, Phys. Rev. Lett. {\bf {69}}, 2411 (1992).

\bibitem{CSY}  A. V. Chubukov, S. Sachdev and J. Ye, Phys. Rev. B {\bf {49}}%
, 11919 (1994).

\bibitem{S}  S. Sachdev, Z. Phys. B {\bf {94}}, 469 (1994).

\bibitem{T98}  H. Chamati, E. S. Pisanova and N. S. Tonchev, Phys. Rev. B 
{\bf {57}}, 5798 (1998).

\bibitem{evst}  R. Evans and J. Stecki, Phys. Rev. B{\it {\ }}{\bf {49}},
8842 (1994).

\bibitem{PE}  A. O. Parry and R. Evans, Physica A {\bf {181}}, 250 (1992).

\bibitem{dietrich}  S. Dietrich, in {\it {Phase Transitions and Critical
Phenomena}}, edited by C. Domb and J. L. Lebowitz (Academic, New York,
1988), Vol. 12.

\bibitem{evans}  R. Evans, in {\it {Liquids at Interfaces}}, Les Houches
Session XLVIII, edited by J. Charvolin, J. Joanny and J. Zinn-Justin
(Elsevier, Amsterdam, 1990), p. 3.

\bibitem{krdi}  M. Krech and S. Dietrich, Phys. Rev. A{\it {\ }}{\bf {46}},
1886 (1992).

\bibitem{I}  J. O. Indekeu, M. P. Nightingale and W. V. Wang, Phys. Rev. B%
{\it {\ }}{\bf {34}}, 330 (1986).

\bibitem{kl96}  M. Krech and D. P. Landau, Phys. Rev. E {\bf {53}}, 4414
(1996).

\bibitem{k97}  M. Krech, Phys. Rev. E {\bf {56}}, 1642 (1997).

\bibitem{Cardy}  J. L. Cardy, in {\it Phase Transitions and critical
phenomena}, edited by C. Domb and J. L. Lebowitz (Academic, New York, 1987),
Vol. 11.

\bibitem{PHA91}  V. Privman, P. C. Hohenberg, and A. Aharony, in {\it Phase
Transitions and critical phenomena}, edited by C. Domb and J. L. Lebowitz
(Academic, New York, 1991), Vol. 14.

\bibitem{EBR}  T. W. Burkhardt and E. Eisenriegler, Phys. Rev. Lett. {\bf {74%
}}, 3189 (1995); E. Eisenriegler and U. Ritschel, Phys. Rev. B {\bf {51}},
13 717 (1995).

\bibitem{SKT}  H. E. Stanley, Phys. Rev.{\it {\ }}{\bf {176}}, 718 (1968);
M. Kac and C. J. Thompson, Phys. Norveg.{\it {\ }}{\bf {5, }}163 (1971).

\bibitem{BF}  M. Barber and M. Fisher, Annals of Physics {\bf {77}}, 1(1973).

\bibitem{R}  J. Rudnick, in {\it {Finite Size Scaling and Numerical
Simulations of Statistical Systems}}, edited by V. Privman (World
Scientific, Singapore, 1990).

\bibitem{Pathria}  S. Singh and R. K. Pathria, Phys. Rev. B {\bf 31}, 4483
(1985).

\bibitem{history}  S. Singh and R. K. Pathria, Phys. Rev. B {\bf 32}, 4818
(1985); J. G. Brankov, J. Stat. Phys. {\bf 56}, 309 (1989); S. Allen and R.
K. Pathria, Can. J. Phys. 67, 952 (1989); S. Singh and R. K. Pathria, Phys.
Rev. B {\bf 40}, 9238 (1989); J. G. Brankov and N. S. Tonchev, J. Stat.
Phys. {\bf 59}, 1431 (1990); J. G. Brankov and D. M. Danchev, J. Stat. Phys. 
{\bf 71}, 775 (1993); D. M. Danchev, J. Stat. Phys. {\bf 73}, 267 (1993).

\bibitem{DB}  J.G. Brankov and D. M. Danchev, J. Math. Phys.{\it {\ }}{\bf {%
32}}, 2543 (1991).

\bibitem{L81}  L. Lewin, {\it Polylogarithms and associated functions}
(North Holland, Amsterdam, 1981).

\bibitem{S93}  S. Sachdev, Phys. Lett. B {\bf {309}}, 285 (1993).

\bibitem{FP}  M.Fisher and V. Privman, Phys. Rev. B{\it {\ }}{\bf {32}}, 447
(1985).

\bibitem{P2}  R. K. Pathria, Phys. Rev A {\bf 5}, 1451 (1972).

\bibitem{SP86}  S. Singh and R. K. Pathria, Phys. Rev. B {\bf 33}, 672
(1986).

\bibitem{AP93}  S. Allen and R. K. Pathria, J. Phys. A {\bf 26}, 6797 (1993).

\bibitem{history2}  E. Br\'{e}zin, J. Phys. (France) {\bf 43}, 15 (1982); J.
M. Luck, Phys. Rev. B {\bf 31}, 3069 (1985); J. Shapiro and J. Rudnick, J.
Stat. Phys. {\bf 43}, 51 (1986); S. Singh and R. K. Pathria, Phys. Rev. B 
{\bf 36}, 3769 (1987); M. Henkel, J. Phys. A {\bf 21}, L227 (1988); S. Singh
and R. K. Pathria, Phys. Rev. B {\bf 40}, 9238 (1989); S. Allen and R. K.
Pathria, Can. J. Phys. {\bf 67}, 952 (1989); M. Henkel and R. A. Weston, J.
Phys. A {\bf 25}, L207 (1992); S. Allen and R. K. Pathria, Phys. Rev. B {\bf %
50}, 6765 (1994); S. Allen and R. K. Pathria, Phys. Rev. B {\bf 52}, 15925
(1995).

\bibitem{BD}  T. W. Burkhardt and B. Derrida, Phys. Rev. B{\it {\ }}{\bf {32}%
}, 7273 (1985).

\bibitem{BD2}  A. D. Bruce, J. Phys. A {\bf {18}}, L873 (1985); D. P. Landau
and D. Stauffer, J. Physique {\bf {50}}, 509 (1989).

\bibitem{B}  Y. Rouault, J. Baschnagel and K.Binder, J. Statist. Phys. {\bf {%
80}}, 1009 (1995); K. Binder, D. P. Landau and A. M. Ferrenberg, Phys. Rev.
Lett. {\bf {74}}, 298 (1995); K.Binder, R. Evans, D. P. Landau, A. M.
Ferrenberg, Phys. Rev. E {\bf {53}}, 5023 (1996).

\bibitem{FL}  A. M. Ferrenberg and D. P. Landau, Phys. Rev. B {\bf {44}},
5081 (1991).

\bibitem{BSD}  H. Saleur and B. Derrida, J. Physique {\bf {46}}, 1043 (1985).
\end{references}
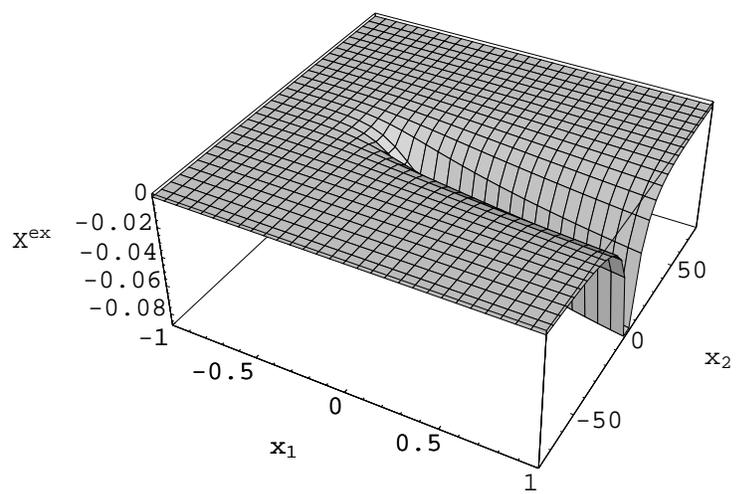
\end{document}